\documentclass[preprint,showpacs,preprintnumbers,amsmath,amssymb]{revtex4}

\usepackage{graphicx}
\usepackage{amsmath}
\usepackage{bm}
\usepackage{latexsym}

\newcommand{\half}{\mbox{$\textstyle \frac{1}{2}$}}
\def\opone{\leavevmode\hbox{\small1\kern-3.8pt\normalsize1}}

\begin{document}

\title{Quantum repeaters with imperfect memories: cost and scalability}

\author{M. Razavi}
\affiliation{Institute for Quantum Computing,}
\affiliation{Department of Electrical and Computer Engineering,}
\email{mrazavi@iqc.ca}
\author{M. Piani}
\author{N. L\"{u}tkenhaus}
\affiliation{Institute for Quantum Computing,}
\affiliation{Department of
Physics and Astronomy, \\University of Waterloo, Waterloo, ON,
Canada  N2L 3G1}

\begin{abstract}
Memory dephasing and its impact on the rate of entanglement generation in quantum repeaters is addressed. For systems that rely on probabilistic schemes for entanglement distribution and connection, we estimate the maximum achievable rate per employed memory for our optimized partial nesting protocol, when a large number of memories are being used in each node. The above rate scales polynomially with distance, $L$, if quantum memories with infinitely long coherence times are available or if we employ a fully fault-tolerant scheme. For memories with finite coherence times and no fault-tolerant protection, the above rate optimistically degrades {\em exponentially} in $\sqrt{L}$, regardless of the employed purification scheme. It decays, at best, exponentially in $L$ if no purification is used.
\end{abstract}
\maketitle

\section{Introduction}
Quantum repeaters enable entanglement distribution between remote parties by relying on a network of quantum memory units \cite{briegel98a, duan01a, razavi06a, qrepeaters, sangouard07a, DLCZvariants, hartmann07a, collins07a, razavi08a, SPIE}. In their seminal paper \cite{briegel98a}, Briegel et al. demonstrate how to distribute entanglement over arbitrarily long distances using ideal quantum memories as well as highly efficient quantum gates for entanglement purification \cite{bennett96a,bennett96b,dur03a}. The resources needed in such a scenario will then only grow polynomially with distance for a fixed desired fidelity for the final entangled state. Further studies have shown that so long as the coherence time $\tau_c$ of the memories is much longer than the transmission delay $L/c$, where $L$ is the distance between the two parties and $c$ is the speed of light in the channel, the above assertion still holds \cite{hartmann07a,collins07a}. It is not clear, however, how the required resources scale in the limit of large distances when $\tau_c$ is finite. Here, we answer this question by assuming that the only error-correction mechanism used in the system is entanglement purification in conjunction with allowing for error-free, but probabilistic, gates to be used instead of erroneous deterministic ones. We find that, even under optimistic assumptions, the system cost explodes as a power of $\exp[\sqrt{(L/c)/ \tau_c}]$ for $\tau_c \ll L/c$, unless we employ a fault-tolerant scheme to remedy the memory decay \cite{hartmann07a, Jiang09a}.
% Whereas, in principle, it is possible to employ fault-tolerant schemes to get around the memory decay problem \cite{hartmann07a}, here . Building such a fault-tolerant system, however, may be as challenging as that of constructing a fully operational quantum computer. Quantum repeaters In this paper, we look instead at the role of different purification schemes and find that, even under optimistic assumptions, the system cost explodes as a power of $\exp[\sqrt{(L/c)/ \tau_c}]$ for $\tau_c \ll L/c$ unless we employ a fault-tolerant scheme to remedy the memory decay. %Asymptotically, this is still an improvement over 

In order to quantitatively address the cost factor in quantum repeaters, we look at the generation rate of maximally entangled states {\em per employed memory} in the system. We employ proper entanglement measures \cite{horodecki00a}, instead of merely looking at the fidelity, to find this rate. We obtain this rate, in the steady state, assuming that the resources in our system are being successively used, according to a proper protocol, to create entangled states. Such a rate-over-cost measure is useful for applications in quantum key distribution (QKD) \cite{ekert91a}, where the generation rate of secure key bits is proportional to the rate of entanglement generation. Moreover, it provides us with a fair and practical measure for comparing different quantum repeater setups and their contrast with alternative schemes for entanglement distribution that do not rely on using quantum memories, such as quantum relay structures \cite{deriedmatten04a} or the direct transmission of entangled photons. In the latter cases, the rate will decay exponentially with distance as a result of loss in the channel.

Memory decay is one of the most challenging problems in quantum repeater technology \cite{memtech}. Its deteriorating effect, however, has not yet fully scrutinized. In \cite{hartmann07a}, authors study the role of memory errors in quantum repeaters, but they treat the required initial entanglement as a given resource. This approach cannot fully capture the memory decay problem as it neglects to account for the corresponding waiting times during initial entanglement distribution. Entanglement distribution, regardless of the employed scheme, is a probabilistic process, mainly because of its dealing with the loss in the channel, and therefore, the time required to entangle two memories is a random variable. Collins et al. \cite{collins07a} consider this probabilistic nature in a multiple-memory configuration, and report a numerical-analytical rate analysis that includes the effects of memory decay. They model the memory decay by associating a lifetime window---within which the stored entanglement is unaffected and beyond which it is destroyed---to each memory. This simple model is not, however, sufficiently realistic to properly account for the effect of memory errors on the rate, especially in the regime of short coherence times. 

In this paper, we introduce a partial nesting protocol, inspired by the blind repeater protocol proposed in \cite{hartmann07a}, and find the throughput in the two cases of perfect and imperfect memories, with or without purification, for a generic quantum repeater system that relies on probabilistic schemes for initial entanglement distribution as well as for its entanglement connection. Such probabilistic architectures for quantum repeaters \cite{duan01a,sangouard07a, DLCZvariants} can adapt themselves to post-selection-based self-purification mechanisms, which---in the absence of memory errors---will allow them to create high-fidelity entangled states over moderately long distances without relying on high-quality quantum gates \cite{razavi06a}. Such an advantage comes at the price of their achieving relatively low entanglement generation rates owing to the probabilistic nature of their operation. They are nevertheless attractive options for QKD systems whose security over long distances can be guaranteed by a combination of such quantum repeater links and sparsely located trusted nodes. To account for the above practical issue, in Sec. II, we describe the probabilistic framework of our multi-memory quantum repeater system. In Sec. III, we present our rate analysis for different scenarios. Numerical results will be presented in Sec. IV, and we conclude the paper in Sec. V.

\section{System Description}

\begin{figure}
\centering
%\begin{tabular}{cc}
%\begin{minipage}{3.0in}
\centering
\includegraphics [width=\linewidth]{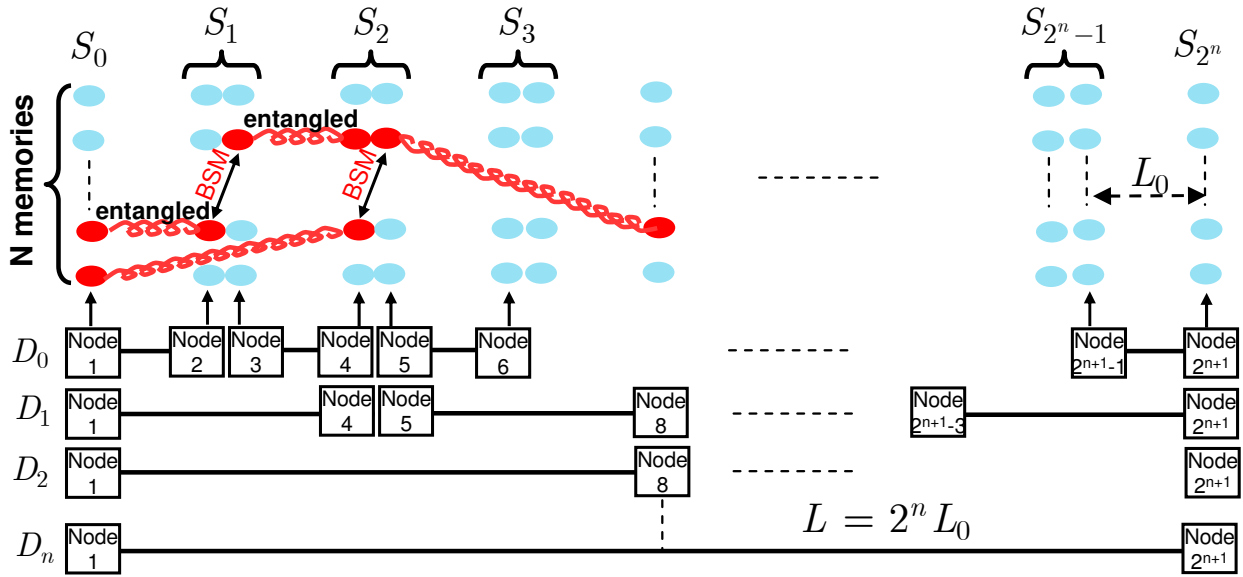}
%\vspace{.2in}
%\parbox{3.0in}  { % 
\caption{
\label{fig1} 
(Color online) A quantum repeater with multiple quantum memories per node. At each round, we first employ an entanglement distribution protocol to entangle any unentangled memory pairs over $D_0$ links. It takes at least $T_{\rm ED} = L_0/c$ to learn about the success/failure of these attempts, which succeed with probability $P_S$, hence $T_{\rm ED}$ is the shortest period at which any protocol can run. At any such cycle, we also match up entangled pairs at different stations, according to our partial nesting protocol (see the text), to perform Bell-state measurements (BSMs), which succeed with probability $P_M$.} %}
%\end{minipage}
\end{figure}

In our analysis, we study a quantum repeater link of length $L$ with $2^n$ sublinks of length $L_0$, as shown schematically in Fig.~\ref{fig1}, with $n$ being the nesting level of the system. At the end of each sublink there is a bank of $N$ quantum memories. We schematically refer to the banks in Fig.~\ref{fig1} by nodes 1 to $2^{n+1}$. Nodes $2i$ and $2i+1$, for $i=1,\ldots,2^n-1$, form physical stations $S_i$ at which Bell-state measurements (BSMs) take place. For future reference, we also denote the family of sublinks of length $2^k L_0$, corresponding to nesting level $k = 0, \ldots, n$, that connect nodes $2^{k+1}(i-1)+1$ and $2^{k+1}i$, for $i=1, \ldots, 2^{n-k}$, by $D_k$, and the set of all stations corresponding to nesting level $k>0$ by $S^{(k)} \equiv \{S_{2^{k-1}(2l-1)}, l=1, \ldots, 2^{n-k}\}$.

The probabilistic framework that we use for our rate analysis is as follows. We denote the probability of success for our employed entanglement distribution scheme over distance $L_0$ by $P_S$. The implicit assumption here is that our entanglement distribution scheme is heralding, i.e., there is a mechanism by which we can verify or become informed of whether our entangling attempt has failed or succeeded \cite{duan01a}. We also assume that the whole process of entanglement distribution from the time that it starts until we learn about its result takes $T_{\rm ED} \equiv L_0/c$. This is the fundamental time period at which we can attempt successively to entangle two specific memories at distance $L_0$. We further assume that, in the absence of memory errors, the memories $A$ and $B$, upon a successful entangling attempt, are in the maximally entangled state $|\psi^+\rangle_{AB}$, where $|\psi^\pm \rangle_{AB} = (|01\rangle_{AB} \pm |10\rangle_{AB})/ \sqrt{2}$ and $|j\rangle_{K}$ is the logical qubit $j=0,1$ for memory $K=A,B$. Similarly, we associate a success probability $P_M < 1$ to each BSM. This probability, in principle, depends on the state of the memories before performing the BSM. In our optimistic analysis, we assume $P_M$ is the maximum probability of obtaining a conclusive result for a given BSM module. For instance, if we rely on a specific pattern of photodetection events for performing a BSM, the inefficiency of photodetectors may result in inconclusive outcomes, in which case we consider the BSM has failed. %A BSM on two entangled states, in general, may reduce the fidelity of the resulting entangled state. In the discussion that follows, however, we first focus on the system rate behavior in connection with its probabilistic components and later discuss the role of errors.

The probabilistic nature of our measurement modules imposes stringent conditions on how a quantum repeater protocol may run. In the nested purification protocol (NPP) proposed in \cite{briegel98a}, in order to extend the existing entanglement on $D_{k-1}$ links to $D_k$ links, we must perform BSMs at stations $S^{(k)}$, for $k=1,\dots, n$. Suppose that we have access to deterministic gates for local operations and measurements, and that no purification is required. Then, after establishing entanglement over all $D_0$ links, one can potentially perform all required BSMs simultaneously at all stations. In our probabilistic setup, however, there is no guarantee that the BSM at one nesting level provides us with entangled states required for the next nesting level. More generally, in a multiple-memory configuration, it is important to know on which pair of memories a BSM can be performed, and for that one needs to communicate between the nodes. This is also the case for the NPP if the employed purification scheme is probabilistic \cite{bennett96a}. 

The above observation requires us to distinguish between two types of BSMs that one can perform. In our protocol, we consider a BSM at nesting level $k=1, \ldots, n$ to be ``informed'' if it is performed on a pair of memories in nodes $2i$ and $2i+1$ {\em known} to station $S_i \in S^{(k)}$ to be entangled with one of the memories in the relevant distant locations. %, respectively, nodes $2i - 2^{k-1}+1$ and $2i + 2^{k-1}$. 
A BSM is called ``blind'' if the above condition does not necessarily hold \cite{hartmann07a}. %To be able to perform informed BSMs, the success/failure of BSMs must be reported to the respective remote nodes either for release of their memories or for use in further entanglement swapping steps. That introduces a transmission delay in our system, which, for the $k$th nesting level, is given by $T_k \equiv 2^{k-1} T_{\rm ED}$.

\begin{figure}
\centering
%\begin{tabular}{cc}
%\begin{minipage}{3.0in}
\includegraphics [width=\linewidth]{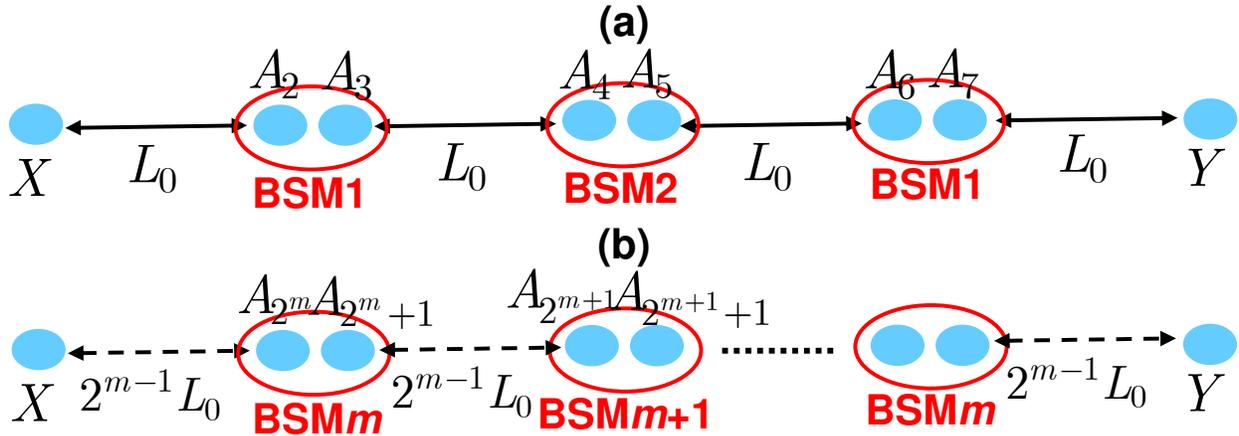}
%\vspace{.2in}
%\parbox{3.0in}  { % 
\caption{
\label{mPNP} 
(Color online) (a) A quantum repeater link with nesting level $n=2$. Here, we have only shown memories $A_2, \ldots, A_7$ that contribute to entanglement generation between memories $X$ and $Y$. In the 2-PNP, all measurements are informed so if measurements labeled by BSM1 occur at time $t_1$, BSM2 will occur at time $t_1 + T_1$. In the 1-PNP, however, all measurements happen at time $t_1$. In such a case, BSM2 will be a blind BSM. (b) A snapshot of memories involved in creating entanglement between $X$ and $Y$ in the general case of $m$-PNP. The assumption here is that we know that $X$ and $A_{2^m}$, $A_{2^m+1}$ and $A_{2^{m+1}}$, and so on are entangled because all previous measurements have been informed. The remaining measurements in $2^n/2^{m-1} -1$ stations shown above all occur at the same time. Among these BSMs, those measurements that correspond to nesting level $m$, labeled by BSM$m$ are informed, and the rest are blind.  } %}
%\end{minipage}
\end{figure}

With the above considerations, our $m$th-order partial nesting protocol ($m$-PNP), for $m=1, \ldots, n$, proceeds as follows. The $m$-PNP is a cyclic protocol with period $T_{\rm ED}$. At each $T_{\rm ED}$-long cycle, we first attempt to entangle unused memories of $D_0$ links. We learn about the success/failure of these attempts by the next cycle. At each cycle, intermediate stations also match up as many memories as possible in their two banks for performing BSMs. In our protocol, we perform {\em informed} BSMs at stations $S^{(k)}$, for $k=1,\ldots, m$, on memories known to be entangled over $D_{k-1}$, and perform {\em blind} BSMs on memories known to be entangled over $D_{m-1}$, but not necessarily over $D_{k-1}$, at stations $S^{(k)}$, for $k=m+1 ,\ldots, n$; see Fig.~\ref{mPNP}. These blind BSMs are performed as soon as such measurements can be done. Purification may or may not be used over the first $m$ nesting levels. After each measurement, or after creating entanglement over distance $L$, the involved memories will be released to be used again in the process of entanglement distribution. The required time for quantum measurement/processing is considered to be negligible throughout the paper. 

The key to the polynomial scaling of the rate in quantum repeaters is in performing {\em informed}, rather than blind, BSMs. That can better be understood by comparing the special case of $m=n$, which corresponds to the probabilistic NPP, with quantum relays \cite{deriedmatten04a}. In quantum relays, one attempts to establish entanglement over $D_0$ and subsequently performs BSMs at all stations without learning about the success/failure of previous entangling attempts. Quantum relays can, in principle, be run at a rate faster than $1/T_{\rm ED}$ and may not need to use quantum memories, but their overall rate of entanglement generation will be proportional to $P_S^{2^n} P_M^{2^n-1} \propto (P_S P_M)^{L/L_0}$, which, for $P_S P_M <1 $, is exponentially decaying with $L$. This is in contrast with the probabilistic NPP, in which, as we will show, for ideal quantum memories, the rate scales polynomially with $L$ as $P_S P_M^n \propto (L/L_0)^{\log_2 P_M}$.

Informed BSMs, however, require us to wait for classical signals. Even if we neglect the operation time for local gates/measurements, as we do throughout the paper, it still takes $T_{k} \equiv 2^{k-1} T_{\rm ED}$, for $k=1,\ldots, n$, to transfer information from a station in $S^{(k)}$ to its closest station in $S^{(k+1)}$. This implies that there is a minimum delay of $T_{\rm ED} + \sum_{k=1}^{m-1} T_k = T_{m}$, in the $m$-PNP, between our initial entangling attempts and the time that we can perform informed BSMs at the $m$th nesting level, during which memory decay is in effect. The memory decay will effectively reduce the rate of entanglement generation. An optimum rate may then be obtained by a combination of informed BSMs in the first few nesting levels and blind BSMs in the remaining final nesting levels.

To minimize the delay incurred by classical communication and the probabilistic nature of entanglement distribution/connection, we have shown that we must use a large number of memories per each node of our quantum repeater \cite{razavi08a,SPIE}. To find an optimistic estimate of the rate, we then assume that $N \gg 1$; the rate behavior for a finite number of memories has been addressed in \cite{collins07a, razavi08a, SPIE}.

%The requirement for back-and-forth classical communication between nodes of a quantum repeater imposes stringent conditions on the coherence time needed for the employed memories. Because of relying on probabilistic schemes for entanglement distribution and connection, some steps must be repeated upon failure. Even if we assume that all required tasks succeed in their first attempts, it still takes $T_k \equiv 2^{k-1} T_{\rm ED}$, for $0 < k \leq n$, to transfer information from a station in $S^{(k)}$ to its closest station in $S^{(k+1)}$. That implies that there is a minimum delay of $\sum_{k=0}^{m-1} T_k = T_m$, in the $m$th-order partial nesting protocol, between our initial entangling attempt and the last set of BSMs. This sets a benchmark for the required coherence time for the memories employed in quantum repeaters, and can only be met if we use a large stack of memories per node in parallel. Throughout the Letter, we then assume $N \gg 1$; the rate behavior for a finite number of memories per node has partly been considered in \cite{collins07a, razavi08a}, and will be addressed more thoroughly in a separate publication.

%which uses a combination of informed and blind BSMs to accommodate the highest achievable rate for different coherence times.

\section{Rate Analysis}
In this section we obtain optimistic estimates for the normalized rate, the rate of entanglement generation per employed memory, in the two cases of perfect and imperfect memories. In the latter case, the only source of error considered is memory dephasing, whose effect on the rate is studied in conjunction with different types of purification.

\subsection{Ideal memories} 
Let us first calculate the rate in the case of ideal quantum memories, i.e., when there is no source of error, and, therefore, no need for purification in the system. In the $m$-PNP, starting with no entanglement at time zero, and for $N \gg 1$, the average number of entangled pairs over $D_n$ that we create at time $T_{m}$, is given by $N_m^{(n)} \equiv N P_S P_M^{(2^{n-m+1}+m-2)}$. Here, $N P_S$ is the average number of entangled pairs over each $D_0$ link at time $T_{\rm ED}$, $N P_S P_M^{m-1}$ is the average number of entangled pairs over each $D_{m-1}$ link after performing $m-1$ levels of informed BSMs, and $P_M^{(2^{n-m+1}-1)}$ accounts for the remaining BSMs, where $2^{n-m}$ of which are informed BSMs for the $m$th nesting level and the rest are blind; see Fig.~\ref{mPNP}(b). At time $T_{\rm ED}$, there are on average $N - N P_S$ unused memory pairs in each $D_0$ link, which can be used to create new entangled pairs. We overestimate $N - N P_S$ by $N$, and repeat the above argument to obtain the same value $N_m^{(n)}$ for the average number of entangled pairs created over distance $L$ at time $T_{\rm ED} + T_{m}$. By reusing the same argument for each cycle, in the steady state, $N_m^{(n)}$ gives an optimistic estimate of the average number of entangled pairs created per cycle. That implies that the steady-state rate of entanglement generation per each of $N 2^{n+1}$ {\em ideal} memories used in the $m$-PNP, $m=1,\cdots,n$, is approximated by
\begin{equation}
Q_m^{(n)} \equiv \frac{N_m^{(n)}}{T_{\rm ED} N 2^{n+1}} 
 = \frac{P_S P_M^{(2^{n-m+1}+m-2)}}{2L/c}.
\end{equation}
Here, $m=n$ represents the special case of the probabilistic NPP, which is the optimum scenario when memories are perfect.

\subsection{Imperfect memories}
Now, let us consider the case of imperfect memories. Here, we model the memory degradation by a dephasing process in which a qubit state $\hat \rho_{A}$, of memory $A$, is mapped, after decaying for a time period $t$, to 
\begin{equation}
\Gamma_t^A (\hat \rho_{A}) = p(t/2) \hat \rho_{A} + [1-p(t/2)]\hat Z_A \hat \rho_{A} \hat Z_A ,
\end{equation}
where $\hat Z_A$ is the Pauli $Z$-operator acting on memory $A$, $p(t)=[1+\exp(- t/\tau_c)]/2$, and $\tau_c$ is the memory coherence time. The above dephasing process maps $\hat \pi_{AB}^\pm \equiv |\psi^{\pm}\rangle_{AB} \langle \psi^{\pm}|$ to the following rank-two Bell-diagonal state
\begin{equation}
\Gamma_t^A \otimes \Gamma_t^B (\hat \pi_{AB}^\pm) =  p(t) \hat \pi_{AB}^\pm + [1-p(t)] \hat \pi_{AB}^\mp \equiv \hat \rho_{AB}^\pm(t),
\end{equation}
with fidelity $p(t)$. We can also show that a BSM on memories $B$ and $C$ of a four-memory system initially in the state $\hat \rho_{AB}^+(t) \otimes \hat \rho_{CD}^+(t)$ leaves $A$ and $D$, up to a local unitary, in the state $\hat \rho_{AD}^+(2t)$. %With this background, in the following, we calculate the effect of dephasing on the total rate.

Using the above model, we estimate $R_m^{(n)}$, the generation rate of maximally entangled states, per employed memory, in the presence of dephasing errors. We first consider the no-purification case, and then allow to use purification, without restricting ourselves to any specific purification protocol. In each case, for any memories $X$ and $Y$ at distance $L$ in the $m$-PNP, we obtain the final state $\hat \rho_{f,XY}$ upon establishment of entanglement between them. We denote by $A_i$ the memory in node $i=2,\ldots,2^{n+1}-1$ that contributes to the successful creation of entanglement between $X$ and $Y$; see Fig.~\ref{mPNP}.

To underestimate the decay effect, we make a further assumption in calculating $\hat \rho_{f,XY}$. In the $m$-PNP, after performing the last set of BSMs, corresponding to nesting levels $m$ and higher, we must in principle wait, before claiming that $X$ and $Y$ are entangled, until we receive the classical data from all middle nodes. The effective state in such a case must then include the decay during this last step of communication. In certain applications such as QKD, however, we can perform the required QKD measurements on $X$ and $Y$ concurrent with the time that blind BSMs are performed, and later confirm whether they are entangled or not. Under the latter assumption, we do not need to account for the classical delays that occur after the last set of BSMs corresponding to nesting levels $m$ and higher. Throughout the rest of this Section, we assume that we are operating under such conditions.

With the above assumptions, we overestimate the number of maximally entangled states that can be obtained out of $N_m^{(n)}$ copies of $\hat \rho_{f,XY}$ by $N_m^{(n)} E_C(\hat \rho_{f,XY})$. For a state $\hat \rho$, $E_C(\hat \rho)$ is the entanglement cost, viz. the ratio $M/K$ for starting with $(\hat \pi^+)^{\otimes M}$ and obtaining $\hat \rho^{\otimes K}$, in the limit of large $K$, using local operations and classical communication (LOCC). It is known that for the asymptotic version $E^\infty(\hat \rho)$ of any ``well-behaved'' entanglement measure, we have $E_D(\hat \rho) \leq E^\infty(\hat \rho) \leq E_C(\hat \rho)$, where $E_D(\hat \rho)$ is the distillable entanglement, i.e., the ratio $M/K$ for distilling $\hat \rho^{\otimes K}$ to $(\hat \pi^+)^{\otimes M}$, in the limit of large $K$, using LOCC \cite{horodecki00a}. Hence, our choice of entanglement cost as our entanglement measure is in line with our other optimistic assumptions. For $\hat \rho_{XY}^+(t)$, we have $E_D(\hat \rho_{XY}^+(t)) = 1-H(p(t))$ and $E_C(\hat \rho_{XY}^+(t)) = H(\half + \sqrt{p(t)[1-p(t)]})$, where $H(p)=-p\log_2 p-(1-p)\log_2(1-p)$ \cite{horodecki00a}. Here, $E_D(t)$ also gives the asymptotic yield for the one-way hashing protocol for purifying $\hat \rho_{XY}^+(t)$ \cite{bennett96b}.
% and our asymptotic approach N-->infty and asymptotic entanglement measures
\subsubsection{Without Purification}
Let us first consider the case where no purification is employed. Suppose that we attempt to entangle $D_0$ links at time zero. Then by the time that we become informed of the establishment of entanglement, memories have already decayed for a period $T_{\rm ED}$. For instance, $X$ and $A_2$ are in the state $\hat \rho_{XA_2}^+(T_{\rm ED})$. At $T_{\rm ED}=T_1$, we also perform the BSMs corresponding to the first nesting level. That leaves $X$ and $A_4$, right after performing the BSM and up to a local unitary, in $\hat \rho_{XA_4}^+(2T_1)$. It takes $T_1$ to inform $A_4$ of the BSM success, hence at time $T_2=T_{\rm ED} +T_1$, $X$ and $A_4$ are in $\hat \rho_{XA_4}^+(3T_1)$. By continuing the same argument, we can show that at time $T_{m}$,
$\hat \rho_{f,XY} = \hat \rho_{XY}^+(t_m^{(n)})$, 
where $t_m^{(n)} \equiv T_{n+1} + (m-1) T_{n}$, for $m=1,\ldots, n$. By time $T_{m}$, we do not necessarily know whether $X$ and $Y$ are entangled, but as mentioned before, in some applications such as QKD, we can often use $X$ and $Y$ at time $T_m$, and later verify whether they are entangled or not. Under this assumption, our optimistic rate is given by
\begin{equation}
\label{rate_nopur}
R_m^{(n)} \approx Q_m^{(n)} E_C(\hat \rho_{XY}^+( t_m^{(n)})), \quad\mbox{(no purification)}.
\end{equation}
Given that, to the first order, $E_C(\hat \rho_{XY}^+(t)) \propto t \exp(-2t/\tau_c)$ for $t \gg \tau_c$, and that $t_m^{(n)}\geq L/c$, the rate in Eq.~(\ref{rate_nopur}) exponentially decays with distance for a finite value of $\tau_c$. This implies that the partial nesting protocol cannot help us avoid the exponential decay of the normalized rate with distance if we do not use any purification.
 
\subsubsection{With Purification}
Now, let us consider the case when we allow to use purification. In the $m$-PNP, our last chance to purify memories is in the $m$th nesting level. In order to obtain an upper bound on the rate, suppose that at the nesting level $m-1 \geq 1$, we are provided with maximally entangled states $\hat \pi^+$ over $D_{m-2}$ links. (We consider the special case of $m=1$ later.) The BSMs at stations $S^{(m-1)}$, upon success, will leave us, up to a local unitary, in $\hat \pi^+$. There will be, however, a delay of $T_{m-1}$ to inform stations $S^{(m)}$ of the success/failure of the above BSMs, during which the memories decay to $\hat \rho ^+ (T_{m-1})$. In the special case of $m=1$, the minimum delay is given by $T_{\rm ED}$, hence the decayed state will be $\hat \rho ^+ (T_{\rm ED})$. The rate is then optimistically approximated by
\begin{equation}
\label{Eq:rate}
R_m^{(n)} \approx Q_m^{(n)} E_C(\hat \rho_{XY}^+(T_m^{(n)})), \quad\mbox{(with purification)},
\end{equation}
where $T_m^{(n)} = \max (T_{\rm ED},T_{m-1})$. Here, for a fixed value of $n$, $E_C(\hat \rho_{XY}^+(T_m^{(n)}))$ is a decreasing function of $m$, whereas $Q_m^{(n)}$ is an increasing function of $m$. That implies that, for fixed values of $\tau_c$ and $L$, there are optimum values of $m$ and $n$ that maximize the normalized rate, and such a maximum may scale better than exponentially with distance, as we show next.

In order to find the asymptotic behavior of $R_{\rm opt} \equiv \max_{m,n} R_m^{(n)}$ in the limit of large distances, we consider all possible choices that we have for our system parameters $m$ and $n$. The only additional assumption in our analysis is that $P_S \propto \exp(-\alpha L_0)$, where $\alpha$ is inversely proportional to the attenuation length of the channel. This is a reasonable assumption as most proposed entanglement distribution schemes require the transmission of a single photon along the channel \cite{duan01a, sangouard07a, DLCZvariants}. Now let us consider the following cases for the optimum values of $m$ and $n$, respectively, denoted by $m_{\rm opt}$ and $n_{\rm opt}$, all in the limit of large distances: 

\noindent
(1) Suppose $n_{\rm opt}$ is finite, then, because $P_S \propto \exp(-\alpha L/2^n)$, $R_{\rm opt}$ degrades exponentially with $L$.

\noindent
(2) Suppose $n_{\rm opt} \rightarrow \infty$ but $n_{\rm opt} - m_{\rm opt}$ is finite. Then the entanglement cost term in Eq.~(\ref{Eq:rate}), will scale as $\exp[-L/(c \tau_c 2^{n_{\rm opt}-m_{\rm opt}+1})]$, which again represents an exponential decay of the normalized rate with distance.

\noindent
(3) Suppose $n_{\rm opt} \rightarrow \infty$ and $n_{\rm opt}-m_{\rm opt} \rightarrow \infty$, then 
\begin{equation}
\label{Eq:Ropt}
R_{\rm opt} \propto \exp(-\alpha L_0^{\rm opt}) \exp \left( \frac {L \ln(P_M)} {2^{m_{\rm opt}-1}L_0^{\rm opt}} \right) E_C(\hat \rho_{XY}^+(2^{m_{\rm opt}-2} L_0^{\rm opt}/(c\tau_c))),
\end{equation}
where $L_0^{\rm opt} = L/2^{n_{\rm opt}}$. (Note that $R_1^{(n)} \leq R_2^{(n)}$, therefore $T_{m_{\rm opt}}^{(n_{\rm opt})} = 2^{m_{\rm opt}-2} L_0^{\rm opt}/c$.) Now, if $2^{m_{\rm opt}} L_0^{\rm opt}$ is finite, then the entanglement cost term in the above equation will be a constant, whereas the second term, so long as $P_M <1$, decays exponentially with $L$. If $2^{m_{\rm opt}} L_0^{\rm opt} \rightarrow \infty$, we can then replace the entanglement cost term in Eq.~(\ref{Eq:Ropt}) with its asymptotic value to obtain
\begin{equation}
\label{Eq:Ropt2}
R_{\rm opt} \propto \exp(-\alpha L_0^{\rm opt}) \exp \left( \frac {L \ln(P_M)} {2^{m_{\rm opt}-1}L_0^{\rm opt}} \right) \exp(-2^{m_{\rm opt}-1} L_0^{\rm opt}/(c\tau_c)).
\end{equation}
In the above equation, there are two competing terms: the second term that decays exponentially with $L/(2^{m_{\rm opt}-1} L_0^{\rm opt})$ and the third term that decays exponentially with $2^{m_{\rm opt}-1} L_0^{\rm opt}$. The best rate-versus-distance scaling can be achieved when these two terms scale the same with $L$. That can be achieved by assuming $2^{m_{\rm opt}-1} L_0^{\rm opt} = \sqrt{\beta L}$, for some $\beta >0$, in which case both above-mentioned terms decay exponentially with $\sqrt{L}$. After optimizing over $\beta$ and assuming that $m_{\rm opt} \rightarrow \infty$, we find that  
\begin{equation}
\label{Eq:Ropt3}
R_{\rm opt} \propto \exp \left( -2 \sqrt {\frac {L \ln(1/P_M)} {c \tau_c}} \right),
\end{equation}
which is exponentially decaying with $\sqrt{L}$, rather than $L$ as obtained for cases (1) and (2). This is in fact the best rate-versus-distance scaling achievable for our quantum repeater system under our optimistic assumptions. Given that the above optimized rate is obtained in the regime of $m_{\rm opt}, n_{\rm opt} \rightarrow \infty$, we can further manipulate the ratios $R_{m+1}^{(n+1)}/R_{m}^{(n)}$ and $R_{m+1}^{(n)}/R_{m}^{(n)}$ to find that, for large distances, the optimum rate is obtained at 
\begin{equation}
L_0^{\rm opt} \approx 2 \ln P_M^{-1}/\alpha ,
\end{equation}
which is a constant identical to what one obtains for the ideal-memory case, and 
\begin{equation}
2^{m_{\rm opt}} \approx \alpha \sqrt{2 L c \tau_c/ \ln P_M^{-1} } .
\end{equation}
This shows that the optimum rate is obtained when $2^{n_{\rm opt}} \propto L$ and $2^{m_{\rm opt}} \propto \sqrt{L}$. Such an improvement over the no-purification case is associated with the employed purification scheme as we discuss next.

The proposed purification schemes can be divided into one-way and two-way protocols \cite{bennett96b}. In most proposed one-way protocols, it is commonly the case that after performing proper local operations on one side, the measurement results are sent to the other side at which, after further manipulation, the purified states are obtained. In other words, the operation of such schemes is equivalent to that of a quantum code, in which errors occurred during both the processing as well as the storage time may be corrected. In most proposed two-way purification protocols, it is commonly the case that we perform local operations on both sides, and exchange the measurement results between the two parties. The purification process can then be completed only after the receipt of classical data. The problem with these protocols is that the faulty memories will decay during the transmission of classical data. That may result in an exponential decay of rate with distance as detailed below.

To see the effect of a typical two-way purification scheme on the normalized rate, suppose that in order to purify $\hat \rho_{AB}$, an arbitrary two-qubit state between two partners $A$ and $B$ at distance $l$, we apply local operations/measurements on $M$ copies of $\hat \rho_{AB}$ to obtain $K<M$ copies of $\hat \rho_{AB}^{\rm pur}$. Then, we communicate between $A$ and $B$, to verify whether the purification is successful. Upon success we perform another set of local operations, modeled by a quantum operation $\Lambda$, on $(\Gamma_{l/c}^A \otimes \Gamma_{l/c}^B)^{\otimes K}[(\hat \rho_{AB}^{\rm pur})^{{\otimes K}}]$. For many schemes proposed in \cite{bennett96a}, $\Lambda$ is just the identity operator due to a discard-or-keep action, so we further assume that the dephasing process modeled by $\Gamma_{l/c}^A \otimes \Gamma_{l/c}^B$ commutes with $\Lambda$. Under these conditions, we show that the fidelity of memories $A$ and $B$ after purification cannot exceed $p(l/c)$. Suppose $\hat \rho_{AB}^F \equiv \Lambda(\hat \rho_{AB}^{\rm pur})$, where $F = {\rm tr} (\hat \pi^+ \hat \rho_{AB}^F)$ is the fidelity of the state after purification without considering the memory decay. The fidelity of such state after memory decay for a period $t = l/c$ will then satisfy
\begin{eqnarray}
\!\!\!\!\!{\rm tr} [\hat \pi_{AB}^+ \Gamma_t^A \otimes \Gamma_t^B(\hat \rho_{AB}^F)] &=& {\rm tr} [\Gamma_t^A \otimes \Gamma_t^B (\hat \pi_{AB}^+) \hat \rho_{AB}^F] \nonumber \\
&=& {\rm tr} [(p(t) \hat \pi_{AB}^+ + [1-p(t)] \hat \pi_{AB}^- )\hat \rho_{AB}^F] \nonumber \\
&\leq& p(t) F + [1-p(t)] (1-F) \leq p(t),
\end{eqnarray}
where the first equality comes from the cycling property of the trace operator, and equality holds for the first inequality if $\hat \rho_{AB}^F = F \hat \pi_{AB}^+ + (1-F)\hat \pi_{AB}^-$. We used $F \leq 1$ in the last inequality. This implies that even if $\hat \rho_{AB}^{\rm pur} = \hat \pi_{AB}^+$, by the end of the above purification procedure, we ideally end up with $\hat \rho_AB^+(l/c)$. Using such purification schemes in the $m$-PNP will then leave memories $X$ and $A_{2^m}$, at best, in the state $\hat \rho_{X,A_{2^m}}^+(T_m)$, right before performing the last set of BSMs. This is the case too for other similar pairs of memories. After performing the remaining BSMs, we end up with 
$\hat \rho_{f,XY} = \hat \rho_{XY}^+(T_{n+1})$, for which the entanglement cost scales as $\exp[-2L/(c\tau_c)]$, and that proves our claim.

Whereas the two-way purification schemes considered above cannot provide us with the optimum rate-versus-distance scaling shown in Eq.~(\ref{Eq:Ropt3}), there are one-way purification schemes that achieve a similar scaling. For instance, in the $m$-PNP, suppose we perform no purification up until the $m$-th nesting level, at which point we perform a one-way hashing protocol \cite{bennett96b} on entangled states distributed over $D_{m-1}$. These entangled memories, before applying the purification and up to a local unitary, are in the state $\hat \rho^+(t_{m}^{(m)}/2)$. To perform the hashing protocol, we need to perform encoding measurements on one side of each entangled link over $D_{m-1}$ and send the results to the nodes on the other side of the link. Suppose the nodes that receive this encoding information are those at which the $m$th nesting level measurements will occur. For all other nodes, we not only can perform, at time $T_m$, the measurements required for the hashing protocol but also, at the same time, we can perform those blind BSMs, according to our partial nesting protocol, needed for entanglement swapping as well as the required measurements on the end nodes with regard to our QKD application. By this trick, only memories located at $S^{(m)}$ will undergo an additional $T_{m}$-long decay due to the transmission delay before getting purified. The normalized rate after performing the above purification as well as the remaining BSMs is then given by $Q_m^{(n)} E_D(\hat \rho_{XY}^+(t_{m}^{(m)}/2 + T_{m}/2))$, which, for $m=1$ and $L_0 \propto \sqrt{L}$, degrades exponentially with $\sqrt{L}$ similar to Eq.~(\ref{Eq:Ropt3}). Here, we used the fact the asymptotic yield of the one-way hashing protocol for our decayed state is given by the entanglement of distillation $E_D(\hat \rho_{XY}^+(t)) \propto \exp(-2 t /\tau_c)$, for $t \gg \tau_c$.

The above example shows that with even one step of purification we can achieve the optimum scaling achievable by a quantum repeater system that only relies on entanglement purification for mitigating the effect of memory errors. This is a weaker requirement than what needed for fully fault-tolerant schemes, which rely on the frequent use of error correction schemes. Although here we assumed that the gates used in the purification scheme are deterministic and error-free, we can possibly relax this assumption by using probabilistic gates for purification. How the rate decays in this new scenario is a matter of further investigation.

%It is worth mentioning that the above estimate is optimistic for several reasons. First, we assume that the decay process has set in only over the last communication step. Moreover, we assume that our purification scheme is deterministic and error free. Also, our $N \rightarrow \infty$ assumption is in accord with our choice of entanglement cost, as an asymptotic entanglement measure, in our analysis. In addition, it results in the minimum constraints on the coherence time of memories. We further remove constraints on the required coherence time by preserving the memory state only up until the last informed nesting level, and not until the remote parties learn which memory pairs are indeed entangled.  

%Here, $f(m)$ represents the number of levels at which purification is used, and $P_f$ denotes its success probability. As can be seen from Eq.~(\ref{Eq:rate}), the partial nesting protocol may result in a higher throughput than the nested purification protocol by using fewer purification steps, and not necessarily because of its shorter transmission delay. 
\section{Numerical Results}
In this section, we numerically study the effect of different system parameters on the normalized rate. In all graphs presented, we assume $P_S = 0.2 \times 10^{-0.01 L_0}$, which is an appropriate choice for the schemes proposed in \cite{duan01a, sangouard07a}. Furthermore, The underlying physical channel is assumed to be an optical fiber with 0.17$\,$dB/km loss, for which $c=2\times10^8\,$m/s and $1 / \alpha = 50\,$km. For numerical purposes, we have used the normalized rate given by Eqs.~(\ref{rate_nopur}) and (\ref{Eq:rate}), for, respectively, systems without and with purification.  

\begin{figure}
\centering
%\begin{tabular}{cc}
%\begin{minipage}{3.0in}
\centering
\includegraphics [width=\linewidth]{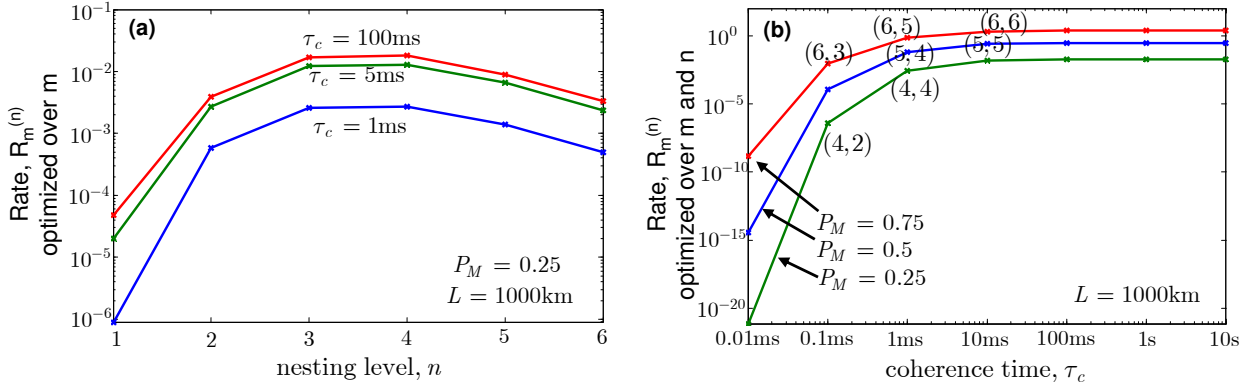}
%\vspace{.2in}
%\parbox{3.0in}  { % 
\caption{
\label{opt_rate_spie} 
(Color online) An optimistic estimate to the generation rate of maximally entangled states per employed memory, $R_m^{(n)}$, for the $m$-th order partial nesting protocol, in the presence of memory dephasing. The implicit assumption is that we employ a sufficiently large number of memories to minimize the waiting times due to classical communication. In (a), we have plotted $R_m^{(n)}$, optimized over $m$,   versus the nesting level, for three different values of coherence time. In (b), we have plotted $R_m^{(n)}$ versus coherence time, optimized over $m$ and $n$. For fixed values of $\tau_c$, the rate scales exponentially in $\sqrt{L}$. The pairs $(n,m)$, on several points on the graph, show the optimum nesting levels, $n$, and the optimum numbers of informed BSMs, $m$, at those points. In both (a) and (b), we assume that our entanglement distribution succeeds with probability $P_S = 0.2 \times 10^{-0.01 L_0}$ and the total distance is $L=1000\,$km, thus $L/c = 5\,$ms in optical fibers.} %}
%\end{minipage}
\end{figure}

Figure~\ref{opt_rate_spie}(a) shows the rate in Eq.~(\ref{Eq:rate}) versus nesting level $n$ for $L=1000\,$km, hence $L/c=5\,$ms. For each value of $n$, we have optimized $m$ to maximize the rate. The optimum values of $m$ turned out to be equal to $n$ for all the marked points in Fig.~\ref{opt_rate_spie}(a). This figure clearly shows the criticality of $L/c$ as a benchmark for the required coherence time. Whereas the rate drops over an order of magnitude when the coherence time is reduced from $5\,$ms to $1\,$ms, there is only a slight improvement in the rate if we use memories with $100\,$ms coherence time. We should bear in mind, however, that these values are only valid in the case of $N \gg 1$. The benefit of using quantum memories with longer coherence time is that we can still achieve the same value of $R_m^{(n)}$ with a lower number of employed memories \cite{SPIE}. Another interesting point in Fig.~\ref{opt_rate_spie}(a) is that the achievable rate for both values of $n=3$ and $n=4$ is about the same. Although, numerically, $n=4$ attains the maximum rate, for practical purposes, it is beneficial to use the less costly architecture corresponding to $n=3$. One last point about this figure is the difference in the slope of rate for values of $n$ smaller than the optimum value of $n$, and those that are larger. It is because for $n < n_{\rm opt}$, loss is the dominant factor, whereas for $n > n_{\rm opt}$ the BSM success probability brings the rate down.

Figure~\ref{opt_rate_spie}(b) shows the optimum rate versus coherence time for three values of $P_M$ at $L=1000\,$km.  On several points on the graph, we have indicated pairs $(n,m)$ that achieve the maximum achievable rate. For $\tau_c \ll L/c$, the partial nesting protocol favors using blind BSMs as they require memories with shorter coherence times. For lower values of $P_M$, one needs to use fewer nesting levels to attain the maximum achievable rate. The optimum values of $n$ in the above cases agree with the relation $L_0^{\rm opt} \approx 2 \ln(P_M^{-1})/\alpha$ found in the previous section.

\begin{figure}
\centering
%\begin{tabular}{cc}
%\begin{minipage}{3.0in}
\centering
\includegraphics [width=.95\linewidth]{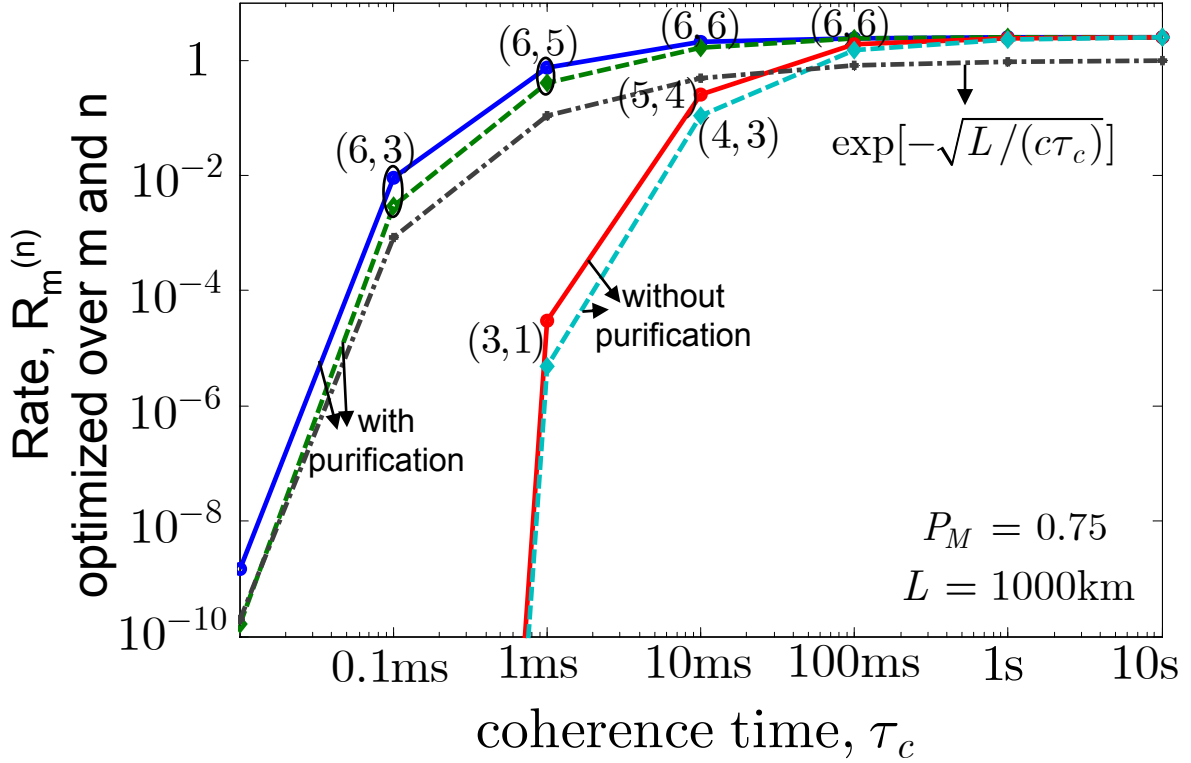}
%\vspace{.2in}
%\parbox{3.0in}  { % 
\caption{
\label{opt_rate} 
(Color online) Estimates for the generation rate of maximally entangled states per employed memory, $R_m^{(n)}$, versus coherence time, optimized over $m$ and $n$, for $P_M = 0.75$, $P_S = 0.2 \times 10^{-0.01 L_0}$, $L=1000\,$km, and $L/c = 5\,$ms. The employed entanglement measures in these graphs are entanglement cost for the solid lines, and entanglement of distillation for the dashed lines. For fixed values of $\tau_c$, the curves scale exponentially in $L$ if no purification is used and exponentially in $\sqrt{L}$ with purification; see dash-dotted line for comparison. The two-tuples $(n,m)$, on several points on the graph, show the optimum nesting levels, $n$, and the optimum numbers of informed BSMs, $m$, at those points.} %}
%\end{minipage}
\end{figure}
 
%Figure~\ref{opt_rate} shows the optimum rate versus coherence time for $L=1000\,$km. Here, we assume $P_S = 0.2 \times 10^{-0.01 L_0}$, which is an appropriate choice for the scheme proposed in \cite{duan01a}, and $P_M = 0.75$. The underlying physical channel is assumed to be an optical fiber for which $c=2\times10^8\,$m/s, hence $L/c=5\,$ms. The solid lines represent the rate when entanglement cost is used as our entanglement measure as in Eqs.~(\ref{rate_nopur}) and (\ref{Eq:rate}). The dashed lines represent the rate when entanglement of distillation is used instead. It can be seen that the difference between the two cases is small. On several points, we have also indicated the pair $(n,m)$ that achieves the maximum achievable rate with and without purification. For $\tau_c \lesssim L/c$, the partial nesting protocol favors using blind BSMs as they require memories with shorter coherence times. Although not shown, for lower values of $P_M$, one needs to use fewer nesting levels to attain the maximum achievable rate.
 
\begin{figure}
\centering
%\begin{tabular}{cc}
%\begin{minipage}{3.0in}
\centering
\includegraphics [width=.95\linewidth]{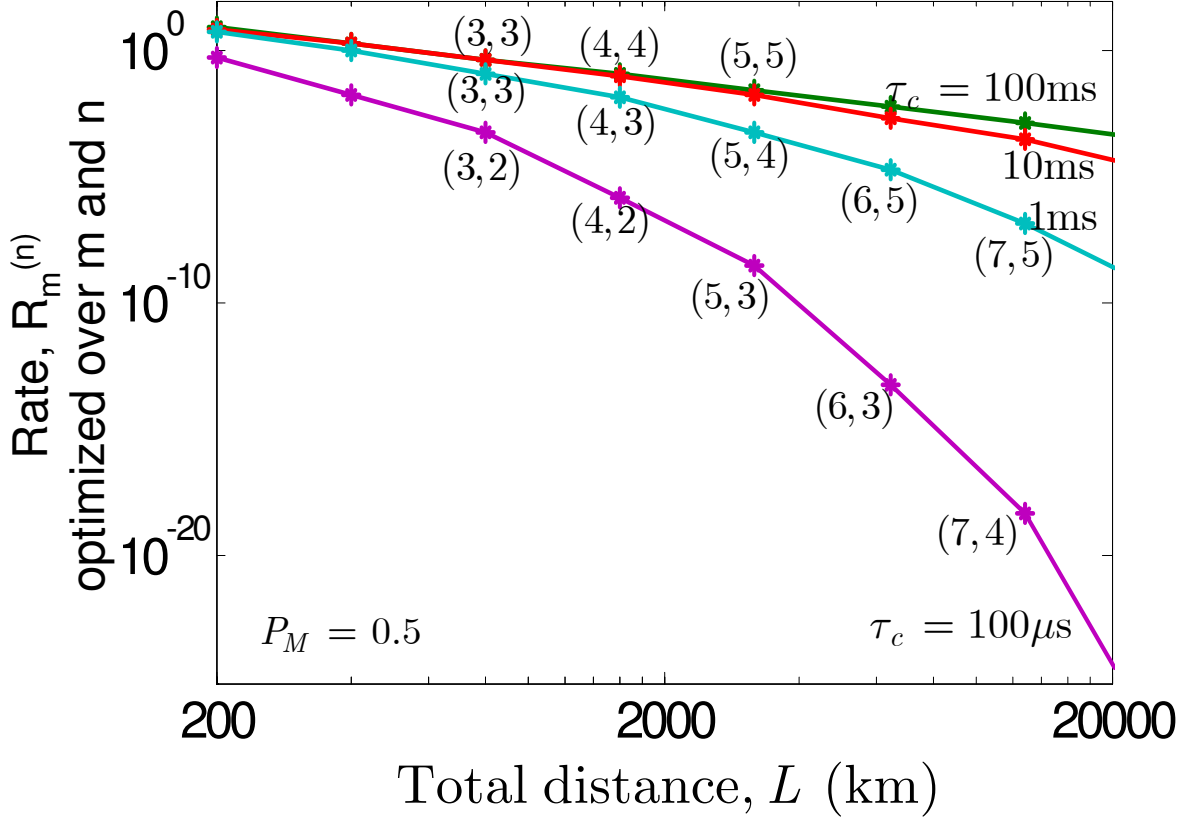}
%\vspace{.2in}
%\parbox{3.0in}  { % 
\caption{
\label{rate_vs_L} 
(Color online) Estimates for the generation rate of maximally entangled states per employed memory, $R_m^{(n)}$, versus total distance $L$, optimized over $m$ and $n$, for different values of coherence time at $P_M = 0.5$ and $P_S = 0.2 \times 10^{-0.01 L_0}$. The employed entanglement measure in these graphs is the entanglement cost and a proper purification protocol is used. It can be seen that, for $\tau_c \gg L/c$, the curves scale polynomially with $L$, whereas, for $\tau_c \ll L/c$, they scale exponentially with  $\sqrt{L}$. The two-tuples $(n,m)$, on several points on the graph, show the optimum nesting levels, $n$, and the optimum numbers of informed BSMs, $m$, at those points. They, respectively, scale with $\log L$ and $\log \sqrt{L}$, for large distances.} %}
%\end{minipage}
\end{figure}

Figure~\ref{opt_rate} demonstrates the rate behavior in both limits of long and short coherence times. The solid lines represent the optimistic normalized rate when entanglement cost is used as our entanglement measure as in Eqs.~(\ref{rate_nopur}) and (\ref{Eq:rate}). The dashed lines represent the achievable normalized rate when entanglement of distillation is used instead. It can be seen that the difference between the two cases is small. It is noticeable that the normalized rate, regardless of the employed entanglement measure or purification scheme, approaches a constant value, $\max_n Q_n^{(n)}$, when $\tau_c \gg L/c$, in accord with what reported in \cite{hartmann07a,collins07a} and Eqs.~(\ref{rate_nopur}) and (\ref{Eq:rate}). In the other unexplored extreme of $L \gg c \tau_c$, however, the situation is much different. In the no-purification case, because, to the first order, $E_C(\hat \rho_{XY}^+(t)) \propto t \exp(-2t/\tau_c)$ for $t \gg \tau_c$, the rate in Eq.~(\ref{rate_nopur}) exponentially decays with distance. Proper use of purification helps, but only a little. As shown by the dash-dotted line in Fig.~\ref{opt_rate}, at the optimum values of $m$ and $n$, $R_m^{(n)}$ in Eq.~(\ref{Eq:rate}) scales as a power of $\exp[-\sqrt{(L/c)/\tau_c}]$. The same effect has been shown in Fig.~\ref{rate_vs_L}, where we have plotted the optimized rate versus $L$ for several values of coherence time. It can be seen that for $\tau_c = 100\,$ms, the logarithm of rate scales linearly with $\log L$, which represent a polynomial scaling of rate with distance. For short coherence times, however, we will observe the exponential decay with $\sqrt{L}$ as discussed before. Another interesting point is the optimum values of $m$ and $n$, which scale, respectively with $\log \sqrt{L}$ and $\log L$.

\section{Conclusions}
In this paper, we studied quantum repeaters that relied on probabilistic schemes for entanglement distribution and connection. We introduced a proper rate-over-cost figure of merit for such systems by looking at the ratio between the rate of entanglement generation---when the system resources are being fully employed to successively create entangled states---and the total number of employed memories. We used entanglement measures to quantify the generation rate of entanglement. We believe that to properly compare different quantum repeater setups, one must study such a normalized rate instead of merely considering the fidelity of generated entangled states at the end of only one round of the protocol. We studied the effect of memory dephasing under the assumption that the only mechanism employed to mitigate errors is purification. We further assumed that the number of memories employed per node is sufficiently large such that, by parallel entangling attempts, we can minimize the unavoidable communication delay between relevant nodes. The memory decay, in principle, can be dealt with using complex fault-tolerant schemes \cite{Jiang09a}, which may not be available in the near future, and has not been considered in our work. We showed that, whereas in the ideal case of infinitely long coherence times the normalized rate scales polynomially with distance, with imperfect quantum memories the rate-over-cost asymptotically scales, at best, as a power of $\exp[- \sqrt{({L/c})/\tau_c}]$, with $\tau_c$ being the memories' coherence time.  This rate behavior is nevertheless superior to the fully exponential decay of rate with distance in quantum relays, which do not use quantum memories, or in systems that do not employ purification but use decaying memories. Our analysis helps us estimate the minimal cost of our quantum communication systems. In the particular case of QKD, to create 1000 secure key bits/s over 1000$\,$km, we may need thousands of memories with coherence times exceeding or on the order of 10$\,$ms. 

\section*{ACKNOWLEDGMENTS} 
We would like to thank W. Munro and K. Nemoto for fruitful discussions. This work was sponsored by QuantumWorks, OCE, and NSERC Discovery Grant.

%\begin{figure}[h]
%\center{\includegraphics[width=.6\linewidth]{logo1.eps}}

%2colourlogo}}

%{\caption{\small QCMC 2008 logo. }}
%\end{figure}

%\bibliography{qit_20080327_Mohsen}

\end{document}